\documentclass{article}

\usepackage{microtype}
\usepackage{graphicx}
\usepackage{subfigure}
\usepackage{booktabs} 
\usepackage{placeins}

\usepackage{hyperref}

\usepackage[accepted]{icml2023}

\usepackage{amsmath}
\usepackage{amssymb}
\usepackage{mathtools}
\usepackage{amsthm}
\usepackage{placeins}
\usepackage{textcomp}

\usepackage[capitalize,noabbrev]{cleveref}

\theoremstyle{plain}

\theoremstyle{definition}

\theoremstyle{remark}

\usepackage[textsize=tiny]{todonotes}

\icmltitlerunning{Challenges of building medical image datasets for development of deep learning software in stroke}

\begin{document}

\twocolumn[
\icmltitle{Challenges of building medical image datasets for development of deep learning software in stroke}



\icmlsetsymbol{equal}{*}

\begin{icmlauthorlist}
\icmlauthor{Alessandro Fontanella}{equal,1}
\icmlauthor{Wenwen Li}{equal,1,3}
\icmlauthor{Grant Mair}{equal,3}
\icmlauthor{Antreas Antoniou}{1}
\icmlauthor{Eleanor Platt}{1}
\icmlauthor{Chloe Martin}{3}
\icmlauthor{Paul Armitage}{4}
\icmlauthor{Emanuele Trucco}{2}
\icmlauthor{Joanna Wardlaw}{3}
\icmlauthor{Amos Storkey}{1}

\end{icmlauthorlist}

\icmlaffiliation{1}{School of Informatics, University of Edinburgh, Edinburgh, UK}
\icmlaffiliation{2}{VAMPIRE project / CVIP, Computing, School of Science and Engineering, University of Dundee, UK}
\icmlaffiliation{3}{Centre for Clinical Brain Sciences, University of Edinburgh, Edinburgh, UK}
\icmlaffiliation{4}{Department of Infection, Immunity and Cardiovascular Disease, The University of Sheffield, Sheffield, UK}

\icmlcorrespondingauthor{Alessandro Fontanella}{A.Fontanella@sms.ed.ac.uk}

\icmlkeywords{Anomaly maps, Counterfactual examples, Diffusion  models, Segmentation masks}

\vskip 0.3in
]



\printAffiliationsAndNotice{\icmlEqualContribution} 

\begin{abstract}
Despite the large amount of brain CT data generated in clinical practice, the availability of CT datasets for deep learning (DL) research is currently limited. Furthermore, the data can be insufficiently or improperly prepared for machine learning and thus lead to spurious and irreproducible analyses. This lack of access to comprehensive and diverse datasets poses a significant challenge for the development of DL algorithms. In this work, we propose a complete semi-automatic pipeline to address the challenges of preparing a clinical brain CT dataset for DL analysis and describe the process of standardising this heterogeneous dataset. Challenges include handling image sets with different orientations (axial, sagittal, coronal), different image types (to view soft tissues or bones) and dimensions, and removing redundant background. The final pipeline was able to process 5,868/10,659 (45\%) CT image datasets. Reasons for rejection include non-axial data (n=1,920), bone reformats (n=687), separated skull base/vault images (n=1,226), and registration failures (n=465). Further format adjustments, including image cropping, resizing and scaling are also needed for DL processing. Of the axial scans that were not localisers, bone reformats or split brains, 5,868/6,333 (93\%) were accepted, while the remaining 465 failed the registration process. Appropriate preparation of medical imaging datasets for DL is a costly and time-intensive process. 

\end{abstract}

\section{Introduction}
\label{sec:introduction}
Deep learning (DL) techniques \cite{lecun2015deep} have risen in popularity and achieved the best performance in many computer-vision benchmarks. At the same time, the interest in DL for medical image analysis is expanding rapidly \cite{esteva2021deep, ting2019artificial, fontanella2023acat, fontanella2023diffusion}. However, the development of successful supervised algorithms, the most common type of deep learning algorithms, requires very large datasets \cite{fontanella2023development}. Due to data privacy concerns, many clinical datasets cannot be made publicly available. With few exceptions, this leads to small, highly curated public medical datasets that limit the applicability of DL methods and do not reflect the variable quality of real clinical images \cite{fontanellaclassification}.

Ideally, DL methods for healthcare should be applicable to unselected, routinely acquired medical images ‘hot off the scanner’. But, in reality, data curation additional to that carried out for clinical studies or real-world care is often required. The challenges of preparing images and related data acquired routinely in clinics, while minimising data loss and maintaining representativeness, are rarely reported and no standardised methodology exists. 

In this study, we present our experience of preparing and standardising data for DL models. Our dataset is composed of brain CT scans collected as part of a large, multicentre clinical trial \cite{ist2012benefits, ist2015association}, from patients with acute ischaemic stroke, and used here as a proxy for routinely acquired clinical data. Indeed, the trial established only minimum requirements, such as whole brain coverage and preferred slice thickness and interval, for the scans to be deemed acceptable. We present a complete data preparation pipeline, where we address issues such as dealing with multiple image series produced by a single visit to the CT scanner, finding only standard axial image data, excluding scans with poor patient positioning and datasets without visible brain tissue (localisers, bone reformats). Additionally, while human readers can ignore factors such as ‘dead space’ outside the head, we needed to further crop, pad, resize and scale images to accommodate the requirement of consistent size and to reduce the influence of extraneous background data on DL performance. 

We aim to provide insights into the complexities involved in preparing clinical CT brain image sets for development of DL algorithms, which we identified in the process of preparing a large pragmatic clinically relevant dataset for DL analysis. By presenting our experience and findings, we intend to offer guidance to researchers working on similar problems in this field and to facilitate the development of DL solutions using real clinical data. Ultimately, we aim to contribute to the standardization of data preparation methodologies in medical image analysis.

\section{Methods}
\subsection{Source Data}
We used CT brain scans from the Third International Stroke Trial (IST-3) \cite{ist2012benefits, ist2015association}, which recruited patients between 2000 and 2011. In particular, IST-3 recruited 3,035 patients with acute ischaemic stroke from 156 centres in 12 countries. 52\% of the patients were female and the median age was 71 years, typical for acute stroke patients. CT scans came from 6 different CT scanner vendors (Siemens, Philips, GE, Hitachi, Toshiba, Picker). 

To ensure that the trial results would generalise to routine clinical practice and to maximise recruitment, the IST-3 imaging criteria stipulated that recruiting centres meet only minimum essential requirements for the CT brain scans as reported previously \cite{ist2012benefits, ist2015association, wardlaw2015protocol} (e.g., whole brain coverage, preferred window level and width, preferred slice thickness and interval). This aimed to minimise delays in clinical care and maximise the relevance of the data collected. Scans were not excluded on the basis of image quality, e.g., if patients moved during scanning, as long as they were deemed satisfactory for diagnosis. The central IST-3 imaging dataset includes scanners from six different manufacturers, different imaging parameters and, as is common in clinical practice, reformatted image sets (all derived from the same raw data) in axial and non-axial orientations, as well as image sets processed with different filters, e.g., for viewing soft tissue and bone. Therefore, although IST-3 images were curated for the trial, they closely resemble data acquired during routine care. 

Ultimately, 95\% of the images collected centrally in IST-3 were from CT scans. All images were pseudonymised using an open-source toolkit for medical imaging de-identification \cite{job2017brain, rodriguez2010open}. Patient names were replaced with individual trial IDs and all identifying data was removed. 

\subsection{Data format}

MRI and CT scans are usually stored in DICOM format (Digital Imaging and Communications in Medicine) \cite{dicom}, an internationally accepted format used by scanner manufacturers and in PACs systems, whereas the NIfTI format (Neuroimaging Informatics Technology Initiative) \cite{nifti} is widely used in neuroimaging research. Both formats combine image files with meta-data (such as details of the patient, scanner, imaging sequence, and how, where and when the image was acquired) in the form of tags or headers. Meta-data can thus include patient identifiable information, sometimes in unexpected locations since many DICOM fields allow the insertion of free text. In a scan, each slice is a 2-dimensional image, in which pixel intensities are the scalar values of the corresponding voxel.

\subsection{Data export and data challenges}

\begin{figure*}[!htb]
\centering
\subfigure[]{
\includegraphics[width=0.5\linewidth]{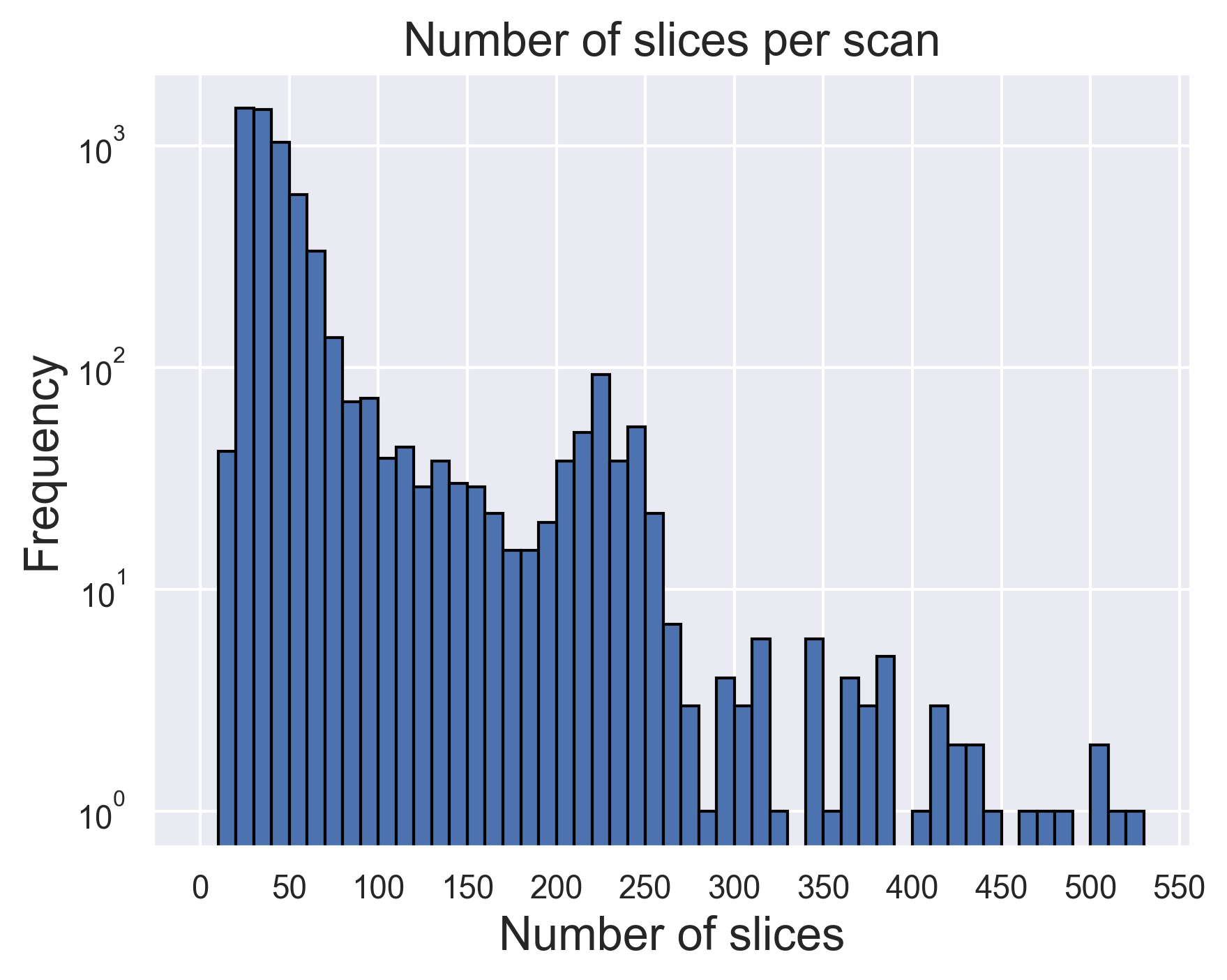}
}


\subfigure[]{
\includegraphics[width=0.45\linewidth]{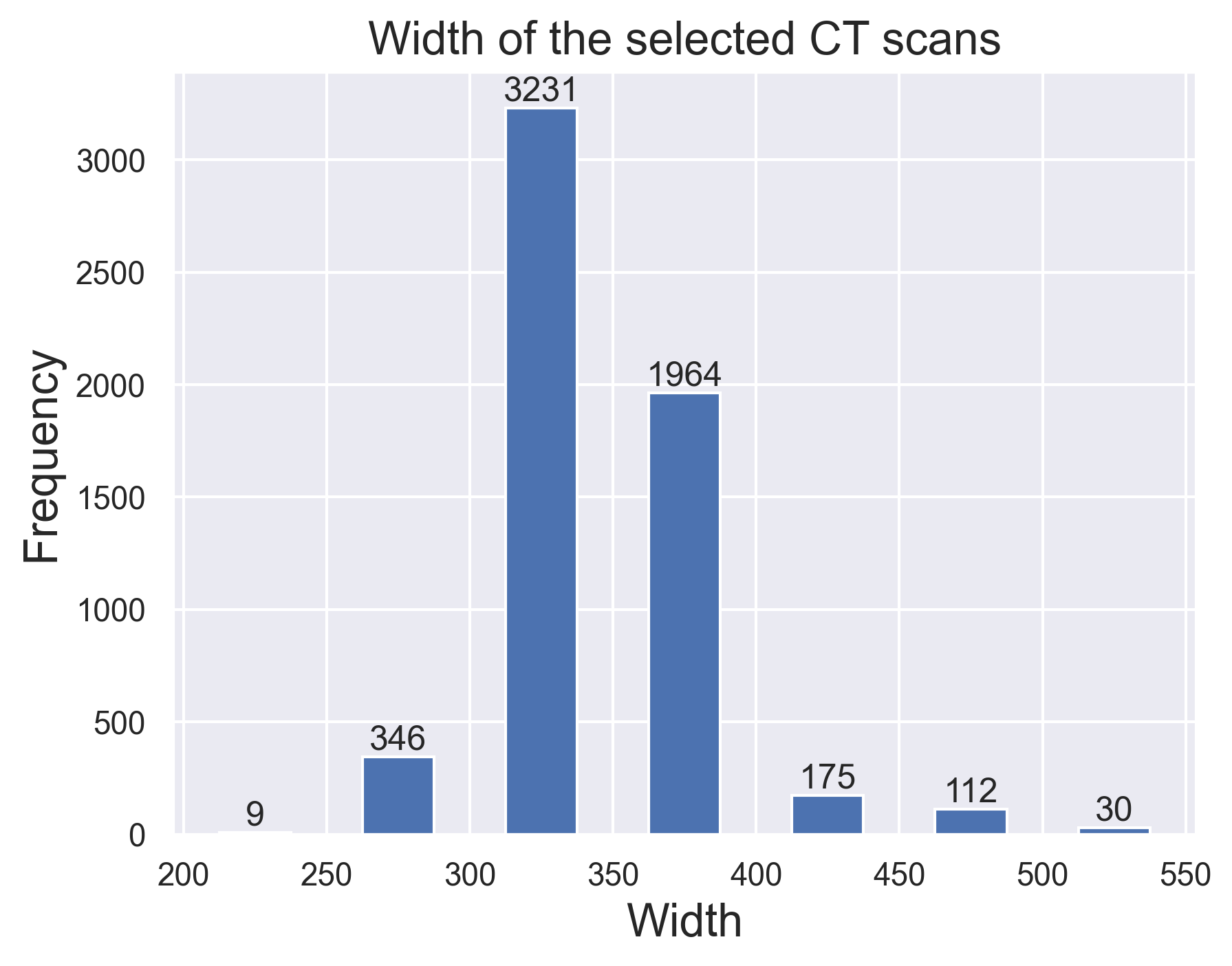}
}
\subfigure[]{
\includegraphics[width=0.45\linewidth]{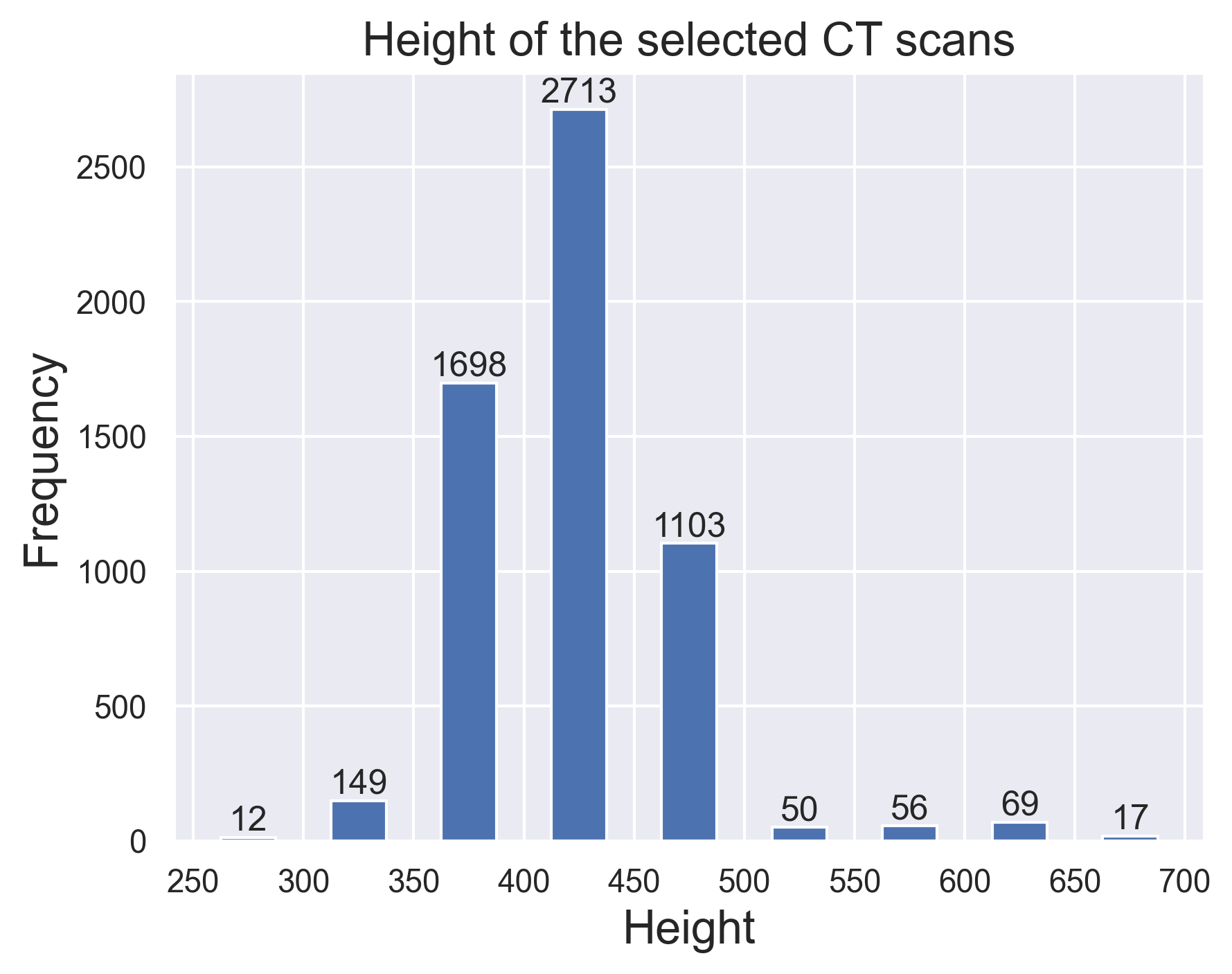}
}

\caption{Distribution of the number of slices per scan in the selected CT data (a), of the width (b) and height (c) of the scans. }
\label{fig:num_slices}
\end{figure*}

All pseudonymised imaging data from IST-3 was first exported from Carestream PACS to a non-proprietary DICOM format using the dcm4che toolkit, which is an open-source implementation of the DICOM standard.  

The exported DICOM dataset was highly variable on initial visual inspection. For example, scans had varying dimensions; the number of slices ranged from 11 to 534 (Figure \ref{fig:num_slices}(a)), the height (Figure \ref{fig:num_slices}(b)) and the width (Figure \ref{fig:num_slices}(c)) from 512 to 800 voxels (from 253 to 699 after removing the background), and from 350 to 650 voxels (from 222 to 512 without background) respectively. Many scan image sets presented different orientations (axial, sagittal or coronal), while others did not include any visible brain tissue; this is for example the case of localisers, which are used to identify the relative anatomical position of a collection of slices within the scan volume. Furthermore, some patients may be ill-positioned during the scan acquisition and the amount of background (bone, extra cranial soft tissue, and room air) surrounding the brain in the CT images is also variable. 

\subsection{Data processing pipeline}

\begin{figure*}[htb]
\centering
\includegraphics[width=0.55\linewidth]{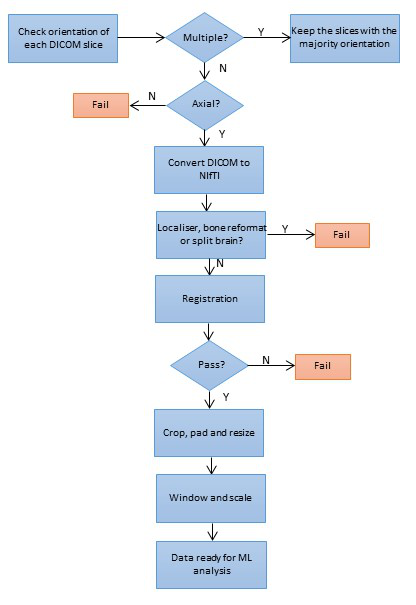}
\caption{Diagram displaying the processing pipeline of a CT scan.}
\label{fig:diag}
\end{figure*}

To handle the heterogeneous nature of our dataset, we developed a data cleaning pipeline to 1) identify axial images, 2) convert DICOM data to NIfTI, 3) remove localisers and scans with separate image sets for skull base and vault, 4) remove bone reformats, 5) remove scans with irredeemable poor patient positioning, 6) crop redundant space around patients, 7) pad/resize image dimensions and 8) scale image brightness (CT Hounsfield Units) for consistency. Brief descriptions of each stage are given below.  Figure \ref{fig:diag} illustrates the data processing pipeline.

1)	Identify axial images

In addition to image sets derived in different viewing orientations (axial, sagittal, coronal), to aid clinical localisation, some scan image sets may contain a mixture of slice orientations, e.g., a single sagittal slice mixed with axial slices. This could lead to errors such as multiple NIfTI files generated while converting from DICOM. The image orientation of individual slices is recognisable based on the Image Orientation DICOM tag.

2)	Data conversion

We used the dcm2niix \cite{li2016first} software for DICOM-to-NIfTI conversion. Clinical CT images can have varying slice thickness, while NIfTI format requires uniform slice thickness. The dcm2niix tool achieves this by interpolation. In the conversion, no details are lost, nor artifacts are introduced.

3)	Remove localisers and scans with separate image sets for skull base and vault

\begin{figure*}[htb]

\centering

\subfigure[Localisers]{
\includegraphics[width=0.18\linewidth]{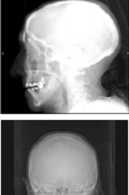}

}
\subfigure[Skull base]{
\includegraphics[width=0.18\linewidth]{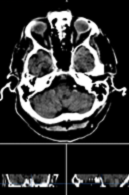}

}
\subfigure[Skull vault]{
\includegraphics[width=0.18\linewidth]{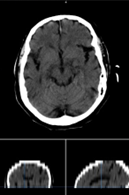}

}
\subfigure[Bone kernel]{
\includegraphics[width=0.18\linewidth]{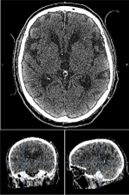}
}
\subfigure[Tissue kernel]{
\includegraphics[width=0.18\linewidth]{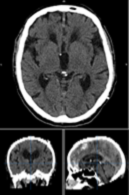}
}

\caption{IST-3 Scan Examples  }
\label{fig.scanexamples}
\end{figure*}

Auxiliary images called localisers are usually acquired to locate the head of the patient within the CT volume and to plan the scan orientation relative to the patient (usually parallel to the anterior skull base for head CT, see Figure \ref{fig.scanexamples}(a)). Such images have no immediate value in DL analysis, since they do not include visible brain tissue. We therefore sought and excluded all localisers using the ImageType DICOM tag. Since this tag may sometimes be blank, we also identified localisers observing the number of slices obtained after conversion to NIfTI. Localisers are converted to independent NIfTI files with only 1 or 2 slices. Therefore, if the ImageType tag is blank, we excluded the NIfTI files with fewer than 3 slices. 

Some older CT scanners (e.g., pre-2011) may produce separate image sets for the inferior third of the head including the posterior fossa and superior two thirds of the head  to allow for greater CT energy through the lower third due to the dense skull base; modern scanners modulate CT energy automatically within a single image set (Figure \ref{fig.scanexamples}(b) and (c)) providing one continuous dataset for the whole head. As the scans consisting of separate inferior and superior parts are structurally incompatible with other scans acquired as a single brain volume, we excluded split scans by manually checking all image sets with fewer than 25 slices in total. We chose this threshold by checking the largest slice number for scans with separate skull base and vault in a random sample of 100 scans with a median slice number under 40 (40 is the median slice number of the whole IST-3 dataset). Scans acquired as split inferior/superior blocks also had different slice thicknesses for the inferior and superior volumes, which could cause registration failures later in the pipeline. 

4)	Remove bone reformats

Raw CT data are filtered to produce images suitable for visual inspection. For CT brain imaging, it is a common clinical practice to routinely produce two image sets, one that is smoothed for optimal viewing of the soft tissues and one that is edge-enhanced to maximise bone details.  The latter has a more granular, noisy texture, and poor discrimination of brain tissue types (cortex and white matter) compared with soft tissue scans and is not used for stroke diagnosis (Figure \ref{fig.scanexamples}(d) and (e)). We excluded bone kernel scans using the ImageType tag.	

5)	Scan registration

Lesions at different brain regions e.g., within the territory supplied by the MCA (middle cerebral artery) or PCA (posterior cerebral artery), may have different characteristics. To enable DL algorithms learning meaningful patterns of lesions at specific brain regions and across different patients, we registered the scans to a common space so that regions are defined consistently across the whole image set.  MRI T1-weighted templates are often used as the standardized coordinate system for registration \cite{kuijf2013registration}. Since in this study we focus on CT scans, we performed MR-CT registration as in previous work \cite{kuijf2013registration, roy2014mr}. In particular, we used the Linear Image Registration Tool \cite{jenkinson2002improved} (FLIRT) from FMRIB Software Library (FSL) to register an MRI template \cite{farrell2009development, royle2013estimated} to each CT scan. The registration tool generates two files for each input NIfTI scan: (1) a registered MR image with the same dimensions as the target CT scan and (2) a transformation matrix specifying the rotation, scaling, skew and translation aligning the MRI template to the target CT.

To preserve the original CT dimensions for downstream analysis, we registered the MRI template to CT scans, yielding a transformation matrix. This means the actual scans used are unchanged; rather we gather registration information for each of them. While registering CT scans to MRI templates would result in CT scans being interpolated to match the dimensions of MRI templates, we opted to record only the transformation to the template reference plane; the interpolation effects of registration transformations can upset downstream analysis. This approach enabled us to identify brain regions on the original CT scans consistently and to obtain the transformation information.

\begin{figure*}[htb]

\centering

\subfigure[]{
\includegraphics[width=0.22\linewidth]{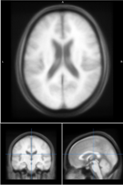}
}
\subfigure[]{
\includegraphics[width=0.22\linewidth]{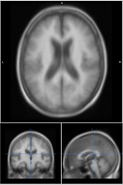}
}

\caption{MRI T1 weighted template for age 65-70 (a) and 75-80 (b).}
\label{fig.mri}
\end{figure*}

In the IST-3 study, 89\% of the patients were over 60 years old and 77\% were over 70.  According to Fillmore et al. \cite{fillmore2015age}, age-specific MRI templates provide less tissue bias in registration than age-inappropriate templates. Therefore, given that the majority of the IST-3 patients were elderly, we chose two T1-weighted age-normalised brain MRI templates \cite{farrell2009development, royle2013estimated}  (Figure \ref{fig.mri}(a) and (b)) previously derived from research into healthy ageing from the brain scans of younger (65 to 70 years, n=54) and older (75 to 80 years, n=25) subjects \cite{farrell2009development}. For MRI-to-CT registration, we used the younger template for IST-3 patients up to 72 years (median age in our dataset) and the older template for IST-3 patients aged 73 years or older. 

\begin{figure*}[t]

\centering

\subfigure[]{
\includegraphics[width=0.25\linewidth]{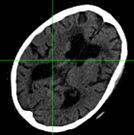}
}
\subfigure[]{
\includegraphics[width=0.25\linewidth]{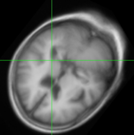}
}

\caption{Example of a registration error. The registration of the slice from a CT scan shown in (a) is off by 180 degrees (b).}
\label{fig.sample}
\end{figure*}

Registration errors may occur: for example, Figure \ref{fig.sample}(a) shows a tilted CT scan, but the registration (Figure \ref{fig.sample}(b)) is off by 180 degrees.  In such cases the brain regions on the target CT could not be identified correctly from the transformation matrix. Such registration errors might be caused, for example, by patients poorly positioned during CT scanning. Since locating brain regions consistently in the original CT scans is crucial to our down-stream analysis, we excluded CT images registered incorrectly. To separate correct from incorrect registrations, we cluster the 3×3 registration transformation matrices for all patients by mapping them to a 3-dimensional space (which preserves 77\% of the variance in the data) using principal component analysis. Then, we perform Gaussian mixture model clustering in the 3-D space. Clusters are then split into valid and invalid by visual assessment, depending on the registration quality of the majority of the scans contained in them. Finally, each registration is accepted or rejected depending on the cluster class it belongs to. 

6)	Cropping redundant background

The larger the field of view (FoV), the smaller the brain appears in the CT image. Hence, depending on the choice of FoV, the raw CT images may contain varying amounts of background which shows no brain, such as regions around the patient and the CT scanner head cradle. To minimise the extent of such uninformative background, we cropped scans to the minimum enclosing rectangle containing the whole head in each slice. To do this, we used the registered CT images. The brain (including the skull) was bounded in a rectangular box (sides parallel to the image edges). The boundary was found by the coordinates of the left-most, right-most, top-most and bottom-most brightest voxel, since skull voxels are the brightest in the scan. Because the registered CT and original CT have the same dimension and orientation, the bounding box obtained from the registered CT was used to crop the original CT.

7)	Image padding and resizing

Most machine learning algorithms and DL frameworks (e.g. Pytorch, Tensorflow) assume that the input dimensions of all the samples are equal. Therefore, to make the height and width consistent across all the scans, we zero-padded or resized each cropped CT image to 500×400 voxels (height×width). This dimension was chosen as more than 95\% of the cropped scans were this size or smaller. In the latter case, dimensions (height and/or width) smaller than the corresponding target size was padded with zeros up to the target size. Images with dimensions larger than the target size were downsized to the target dimensions using the Pytorch interpolate function. This choice minimized interpolation effects.

8)	Scaling image brightness

Each voxel in a CT scan has a numeric Hounsfield unit (HU) \cite{hounsfield1980computed} value indicating CT attenuation, which varies depending on the attention coefficient (i.e. absorption) of radiation within different tissues. These HU values range from -1000 (air) to approximately +3000 (very dense material such as metal). CT scanners are calibrated so that pure water is HU = 0. HU values are displayed as a grayscale image where air is black and very dense materials are white. Since we are only interested in the soft tissues (which comprise a very short section of the total range), we windowed the HU values between 0 and 100. We further scaled the HU values between 0 and 1 by dividing each voxel’s HU value by 100.

\section{Results}

\begin{table}[htb]
\caption{Data lost during the pre-processing.}
\label{table1}

\begin{center}
\begin{tabular}{cc}
\toprule
Reasons for scan exclusion & N\textsuperscript{\underline{o}} of excluded NIfTI files \\
\midrule
Non-axial orientations    & 1920 \\
Localisers & 493\\
Bone reformats     & 687 \\
Separated skull base/vault    & 1226   \\
Poor patient positioning    & 465\\
\midrule
Total      & 4791 \\
\bottomrule
\end{tabular}
\end{center}

\end{table}

From 3,035 patients recruited in IST-3, the vast majority (95\%) of available scans were CT \cite{mair2018effect}, the rest MRI. Patients were recruited by IST-3 during the worldwide transition of medical imaging format from film to digital (DICOM); due to data corruption from long-term DICOM storage prior to the current study, CT data from only 2,578 (85\%) patients (1,247 female and 1,105 male) were exported successfully from our DICOM server. This included 10,659 CT image sets, on average 2.11 per patient. 

The 10,659 CT image sets were highly variable, with non-axial orientations (1,920/10,659, 18\%), localisers (493/10,659, 5\%), bone reformats (687/10,659, 6\%), separated skull base/vault (1,226/10,659, 12\%) or poorly positioned patients whose scans failed registration (465/10,659, 4\%) (Table \ref{table1}). After processing steps 1-5, 5,868 image sets from 2351 patients were selected, each representing a unique CT scan (55\% of those sent to pipeline).

For all 5,868 scans, we cropped any excess background caused by different FoV and standardised inconsistent HU value ranges (the minimum HU value ranged from -32,767 to 0 and the maximum HU value from 255 to 32,767) between 0 and 1. Finally, all 5,868 scans were standardised to the same 500×400 voxel dimensions. 

Our data processing pipeline required on average 2 minutes for processing a scan with 40 slices, the median slice number in IST-3 data, to our DL-ready format. The processing time of each scan can vary up to approximately three-fold depending on the slice number, patient position (for registration), and whether multiple orientations exist for a given CT scan.

Approximately 250 working person-days were spent building this pipeline. This includes the total days of two DL specialists (with limited prior experience handling medical imaging data) and one medical imaging expert spent understanding IST-3 imaging data heterogeneity and exploring previously used methods for processing and standardising CT data, developing and checking the processing pipeline. The heterogeneity of the way scans were represented in DICOM format by different scanners or different institutions created many potential issues affecting small numbers of scans. After investigating each issue, we found that the majority did not affect the process, but in each case this had to be clearly established

\section{Discussion}
The complexity of preparing curated data for AI methods has been addressed by Willemink et al. \cite{willemink2020preparing}, who described fundamental steps for preparing medical images and explained the importance of each step. In particular, they discussed ethical approval, data access and querying, data de-identification, quality control etc, all of which had already been performed on the IST-3 data as part of the original trial. The authors provide guidelines to start the data acquisition and processing, but they do not detail the numerous practical steps required to use clinically relevant CT scans for DL. Muschelli \cite{muschelli2019recommendations}  suggested tools and a pipeline for processing CT data, using Matlab or Python packages to read DICOM data, converting DICOM to NIfTI, choosing appropriate convolution kernels, extracting brain, defacing and registering to a CT template \cite{rorden2012age}, their sample was composed of 30 participants with mean age 61 years.  IST-3 data is substantially larger, more complex, and includes older patients (more representative of clinical practice) with varying FoV, image dimensions and slice numbers, mixed orientations, etc. These issues have not been highlighted previously and we are not aware of papers discussing in detail the processing needed to transform clinical CT data into a format suitable for DL algorithm development.

Our pipeline could be further enhanced by improving the detection of scans with different orientations (axial, sagittal and coronal), better registration between CT and MRI templates (perhaps by using age-normalised CT templates), and by investigating the impact of data quality such as poor patient positioning on the performance of resulting machine learning algorithms. Ultimately, to stratify such cases by quality and include or exclude them accordingly.

Our study is limited to a single dataset which had been through a curation process within the IST-3 trial. However, as discussed above, curation has been deliberately kept to a minimum, and the data set is highly heterogenous, making it well representative of routinely acquired clinical CT scans.

Our pipeline may of course perform differently than reported here with different data and therefore require case-specific adaptations. However, we expect that the pipeline structure and the underlying protocol would remain substantially unchanged. 

Finally, since to our best knowledge no open-source pipeline is available, we were unable to compare our method with other approaches experimentally. 

\section{Conclusion}
While previous studies have described procedures for managing medical imaging data, including acquisition, storage, transfer and anonymisation, many only mention pre-processing in passing and do not provide detailed or complete information on the subject. Data pre-processing was often not the primary focus of these papers, resulting in incomplete discussions of the topic. Our study specifically proposes a pipeline for processing the highly diverse medical data that typically occurs in day-to-day clinical practice, using CT as an example. Our dataset contains CT scans collected over many years with different manufacturer’s scanners from 156 stroke centres in 12 countries. We describe how to address the many issues caused by highly heterogeneous data with variable dimensions, orientation, type, and quality. Our pipeline offers a comprehensive, unified semi-automated solution to standardize clinical data for use in machine learning. We believe that by making our pipeline openly available we can help to bridge the gap between unprocessed clinical data and the refined data required for effective and representative machine learning model training.

\section*{Acknowledgements}
The principal funder was Health Data Research UK (Grant ID:EDIN1). The RS Macdonald Seedcorn Fund supported purchase of GPU hardware. GM is the Stroke Association Edith Murphy Foundation Senior Clinical Lecturer (SA L-SMP 18/1000). JMW  is partially funded by the UK DRI. AF is supported by the United Kingdom Research and Innovation (grant EP/S02431X/1), UKRI Centre for Doctoral Training in Biomedical AI at the University of Edinburgh, School of Informatics.
The funders of this study had no role in the study design, data collection, data analysis, data interpretation, or writing of the report.

\newpage

\bibliography{example_paper}
\bibliographystyle{icml2023}

\newpage
\appendix
\onecolumn


\end{document}